# Citation *proximus*: the role of social and semantic ties in citing behaviour

Diego Kozlowski [1], Carolina Pradier [1], Pierre Benz [1], Natsumi Shokida [1], Jens Peter Andersen[2], Vincent Larivière[1, 3, 4, 5]

[1] École de bibliothéconomie et des sciences de l'information, Université de Montréal, Montréal, QC, Canada.
[2] Danish Centre for Studies in Research and Research Policy, Aarhus University, Aarhus, Denmark
[3] Consortium Érudit, Montréal, QC, Canada.
[4] Observatoire des Sciences et des Technologies, Centre interuniversitaire de recherche sur la science et la technologie, Université du Québec à Montréal, Montréal, QC, Canada.
[5] Department of Science and Innovation-National Research Foundation Centre of Excellence in Scientometrics and Science, Technology and Innovation Policy, Stellenbosch University, Stellenbosch, Western Cape, South Africa.

**Abstract**

Citations are a key indicator of research impact but are shaped by factors beyond intrinsic research quality, including prestige, social networks, and thematic similarity. While the Matthew Effect explains how prestige accumulates and influences citation distributions, our study contextualizes this by showing that other mechanisms also play a crucial role. Analyzing a large dataset of disambiguated authors (N=43,467) and citation linkages (N=264,436) in U.S. economics, we find that close ties in the collaboration network are the strongest predictor of citation, closely followed by thematic similarity between papers. This reinforces the idea that citations are not only a matter of prestige but mostly of social networks and intellectual proximity. Prestige remains important for understanding highly cited papers, but for the majority of citations, proximity—both social and semantic—plays a more significant role. These findings shift attention from extreme cases of highly cited research toward the broader distribution of citations, which shapes career trajectories and the production of knowledge. Recognizing the diverse factors influencing citations is critical for science policy, as this work highlights inequalities that are not based on preferential attachment, but on the role of self-citations, collaborations, and mainstream versus no mainstream research subjects.

# Introduction

In the eyes of many scientists and policy-makers, citations are a key marker of research impact. They have been used—either directly or through proxies, such as the journal Impact Factor—for many decades to assess researchers—both in terms of career progression (McKiernan et al., 2019; Rice et al., 2020) as well as funding allocation (Bol et al., 2018)—and are central components of university rankings (Szluka et al., 2023). However, citation-based indicators have been shown to suffer from several issues, including, such as their lack of comparability across disciplines, their time to accumulate, as well as their skewed distributions, with a minority of papers accounting for the majority of citations (Sugimoto & Larivière, 2018).

Many studies have also acknowledged the multiple factors that affect papers' citations rates, which are not related with their *intrinsic* quality. Among those, the effects associated with markers of prestige—such as authors' characteristics (Beigel et al., 2023; Dworkin et al., 2020; Kozlowski et al., 2022; Sugimoto & Larivière, 2023), institutions (Wapman et al., 2022), or journals' reputation (Larivière & Gingras, 2010; Traag, 2021)—have been heavily studied in bibliometrics, with the Matthew Effect (Merton, 1968)—and its mirror effect, the Matilda effect (Rossiter, 1993)—probably being the most studied (Allison et al., 1982; Dannefer, 1987; DiPrete & Eirich, 2006; Youtie et al., 2013). Those markers of prestige are key drivers of papers' citations, and their cumulative nature shapes the distribution of impact, in which only a few articles and authors receive high numbers of citations, and the majority receive only a handful of citations (Seglen, 1992). Such power-law distributions are coherent with Bourdieu's account of prestige accumulation as a driver of symbolic capital (Bourdieu, 2004), which is unevenly distributed within the scientific field.

While those previous studies highlight the *social* nature of citation links, they do not provide a complete picture of the factors that affect papers' citations. Drawing on a large database of disambiguated authors (N=43,467) and citation linkages (N=264,436), this paper aims to contribute to this literature by analyzing how the social proximity between the citing and cited authors, as well as the semantic ties between the citing and the cited paper, affect papers citation rates.

Social proximity refers to the degree of personal acquaintance or direct social connection between the authors of academic papers. We approximate this phenomena through authors' co-authorship network (Wallace et al., 2012), assuming that the closer two authors are in the network, the more they are likely to know each other (Abbasi et al., 2014; Biscaro & Giupponi, 2014; Uddin et al., 2013; Yan & Ding, 2009). In a citation context, such social ties also encompass authors' self-citations (Bonzi & Snyder, 1991; Costas et al., 2010; Gálvez, 2017; Vincent-Lamarre & Larivière, 2023; Zhao et al., 2018), representing the highest degree of social proximity between citing and cited authors. Other types of social proximity and their effect on citations have only been partially studied (Milard & Tanguy, 2018). Semantic proximity refers to the content similarity between two papers (Ding et al., 2019). While one would expect that citing and cited papers would be thematically related (Frost, 1979), only recently computational methods allowed for large-scale robust comparisons of documents (Kozlowski et al., 2021).

# Results

## Social and semantic distances

When comparing pairs of citing and cited documents with all connected pairs of articles (i.e. pairs of documents which have a non-infinite distance in the collaboration network), notable differences emerge in terms of their social and semantic distances. Figure 1A illustrates the distribution of articles based on their degree of separation within the collaboration network. The general distribution (blue) follows a Gaussian curve centered around six degrees of separation between the authors of two articles. In Economics, this is the most typical distance, which aligns with Milgram's small-world theory (Milgram, 1967). At smaller distances, smaller ego-networks involve fewer authors and, consequently, fewer publications. For example, a distance of 0 indicates that both publications share at least one author, so the share of publications of distance zero is bounded by the previous publications of the authors, which is a small proportion of all the publications in the field.

The distribution of citing-cited pairs (red), however, shows a different pattern. The proportion of self-citations varies significantly between these two sets. While a distance of 0 is negligible among random article pairs, it accounts for nearly 20% of citation pairs. For citing-cited pairs, the distribution is bi-modal: beyond self-citations, the distribution is centered around four degrees of separation, rather than six. Overall, the distribution of citing-cited pairs is shifted towards closer distances and displays a left-skewed shape.

Figure 1B presents the distribution of semantic similarity between articles. Due to computational complexity, to analyze non citing document pairs we used a sample size equivalent to that of the citing-cited pairs. While the absolute values of cosine similarity are not directly interpretable—as they depend on the embedding space used to project the documents—, comparisons between groups remain meaningful (Ait-Saada & Nadif, 2023). The results reveal that non citing document pairs exhibit a more symmetrical and leptokurtic distribution, centered around lower similarity values. In contrast, citing documents pairs show higher semantic similarity.

When analyzing the combined distribution in Figure 1C, we see that the random sample of non citing document pairs shows a more leptokurtic distribution. While self-citations tend to have greater semantic similarity and exhibit higher dispersion, there is no strong correlation between social distances and semantic similarity when examining each group separately. Citing-cited pairs are closer both socially and semantically; however, these two patterns operate independently of each other.

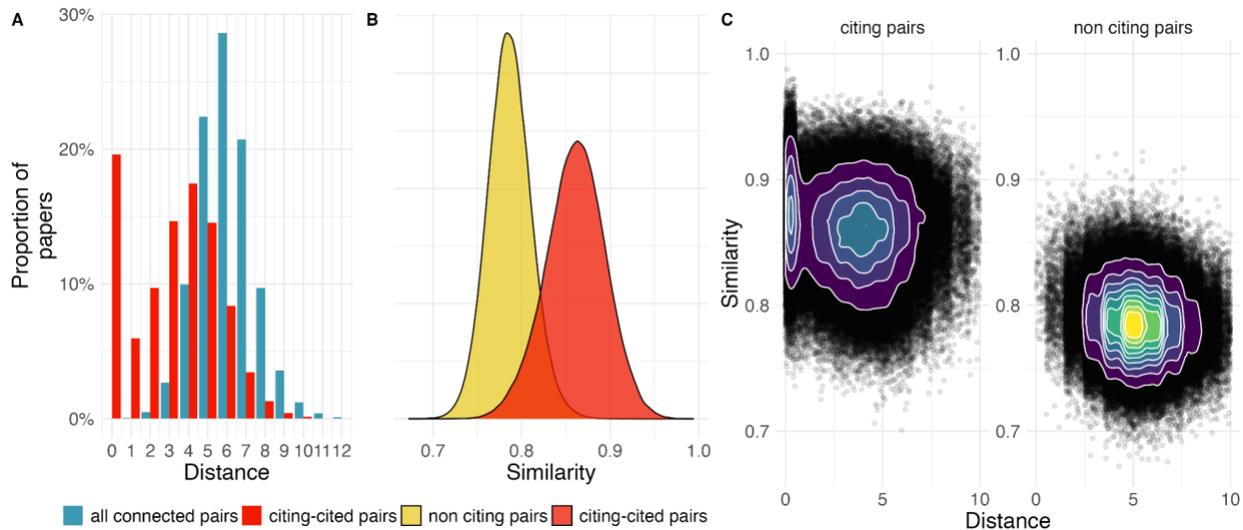

**Figure 1. Distribution shift of citing papers**. A) Distribution of degrees of separation in the collaboration network for the citation pairs and all paper pairs with a non-infinite distance. B) Distribution of cosine similarity of the embedding representation of citing-cited pairs and a random sample of non citing document pairs. C) Combined distribution of distance and similarity for citing-cited pairs and non citing document pairs.

Observing the distribution of distances by deciles of authors' citations (see Fig S1) reveals that prestige significantly influences the distribution. Less cited authors tend to have a larger proportion of self-citations, while highly cited authors gather more citations from documents separated by three to four degrees. This finding is aligned with the Matthew Effect, as the articles of less prestigious authors do not have a cumulative citation advantage from their peers, so they are mostly cited by themselves. Moreover, there is a lower bound in the number of self-citations that an author can receive based on the total number of papers they published, and highly cited authors naturally go beyond this bound. Additionally, due to their more extensive collaboration network, more prestigious authors tend to be more central in the network, with their papers being most frequently at five rather than six degrees of separation of all other papers. This aligns with previous studies that show how centrality in the collaboration network is associated with prestige and accumulated citations (Yan & Ding, 2009). On the other extreme, the least prestigious authors have a larger than average distance to other authors.

When considering other fields (see Fig. S2) a commonality in the main patterns described above can be observed, with a high relevance of self-citation and left skewness among citing-cited pairs. Nevertheless, we can also observe that some smaller fields show very small networks, which might be an indication of the problems of building a self-contained network of co-authorship around small disciplines. On the other hand, semantic similarity does not carry this construction problem and shows very similar patterns across disciplines, and similar degrees of overlap between the distributions of citing and non citing document pairs.

# What drives citations?

The varying distances observed between cited pairs and others raise the question of how these three elements —social distance, semantic proximity, and prestige— influence the decision of citing an article. Figure 2 shows the results of a logistic regression that predicts the probability of a citation between a pair of articles, considering the influence of these three elements, their interactions, and a fixed effect at the cited article level to control for remaining unseen factor, such as the intrinsic quality of each cited article (see Data and Methods).

The results show striking differences between these three explaining factors. Figure 2A shows the AME of each of the independent variables. There, we can see that, by far, the best predictor of a citation is the shared authorship between two documents, that is, a self-citation. If the two documents share an author, there is a 25% greater chance that a citation occurs with respect to a document authored by people at a distance of 6 or more degrees of separation. If the paper was written by a direct co-author, the chances of a citation increase by more than 10%, and this effect decreases as we move further away in the collaboration network. The second most important factor is the semantic similarity between documents. This effect is at the middle point between self-citations and direct collaborations in terms of AME, and relates directly to the role of citations. The fact that citations are more often between semantically related works, after controlling for other social factors, is also a validation of the epistemic function of citations. This is, citing a document is not just a social phenomenon. In this respect, the strength of prestige as a predictor is revealing in the sense that it contextualizes the Matthew effect. Although significantly above 0, the effect of the total citations of the cited authors is marginal with respect to the effect of social and semantic proximity, and only comparable to a distance of 4 degrees of separation. This means that in our model, being the co-author of the co-author of the co-author of the co-author of a paper seems to have a similar influence to being a very highly cited author in the field. It is worth noting that this result is relevant to understand what drives the decision of a single citation link. However, cumulative effects must be considered to understand aggregate results. While the prestige of an author is defined equally across all articles in the field, social and semantic distances are unique to each pair of documents, and therefore prestige will carry greater cumulative effects than these other elements.

The interactions between these three different drivers of citations are also revealing. Figure 2B shows the predicted probability of a citation between two articles for combinations of different values of social distance, semantic similarity, and prestige. If we focus on the interaction between prestige and similarity (left), we can observe that prestige does play a role, but only at similarity levels close to the median. If similarity is significantly below or above the median, the probability of a citation has no relation to prestige. On the other hand, the relation between semantic similarity and social distance (middle) shows a stronger interaction, except in the case of self-citations. For self-citations, the probability of a citation always remains high despite variations in semantic similarity. For all other cases, we observe a diagonal that defines the shift in the probability of citation. Finally, in the relation between social distance and prestige (right), we can see both an absolute gradient for distance, but also the effect of prestige for degrees of separation two and higher. Overall, we observe that the effect of prestige only seems to be relevant for more distant papers, both semantically and socially.

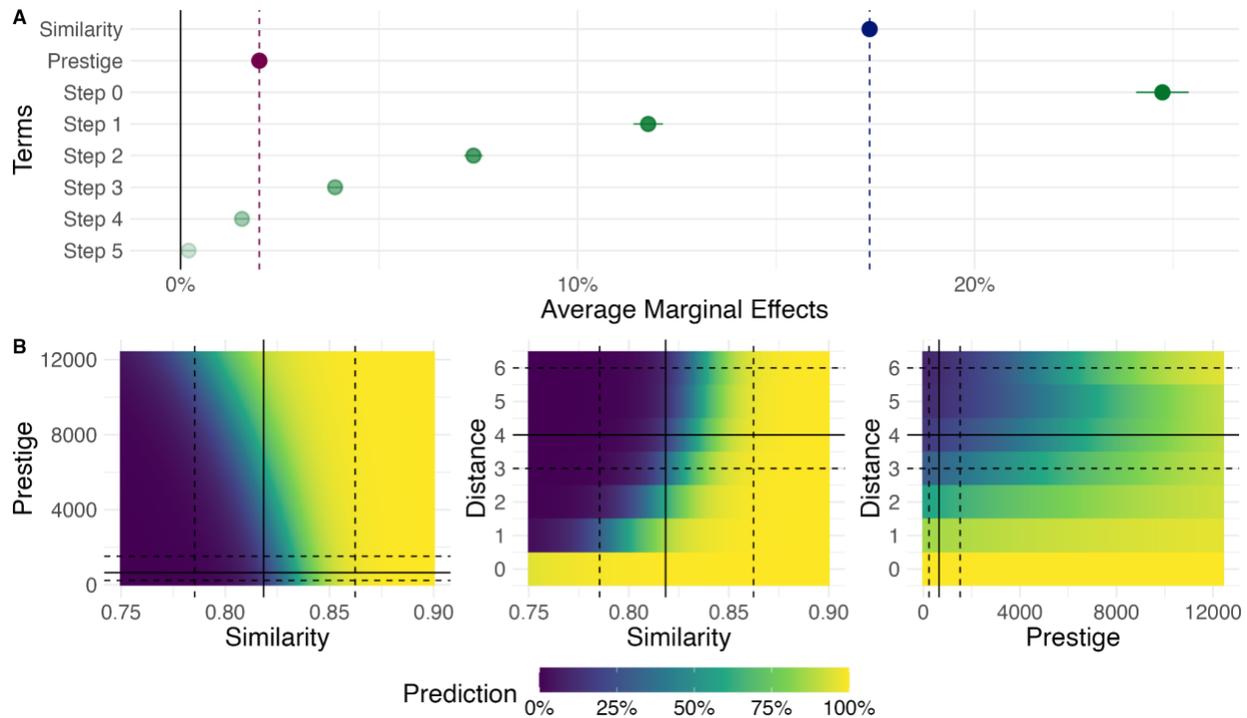

**Figure 2. The effects of social distance, semantic similarity, and prestige on citation links.** A) Average Marginal Effects of cosine similarity between papers, accumulated citations of cited authors and distances on the collaboration network on the existence of a citation link. B) Predicted probabilities at the interaction of independent variables. Solid lines represent the median of each variable, while dashed lines represent the first and third quartile of their distribution

# Discussion

In this study we analyze the effect of social proximity, semantic similarity, and prestige on citations. Our results show that the biggest predictor of a citation is shared authorship. We do not study the causality behind self-citations, but this could be interpreted either as the expression of non-ethical practices or simply the fact that authors know their work more than anyone else and tend to work on similar topics and research lines, which makes their previous work idoneous and accessible for citations (Vincent-Lamarre & Larivière, 2023). The second most important predictor of citations is the semantic similarity between documents. This means that citations do have a meaning beyond social relations. We do not only cite because of a social construct, but because previous work is related to ours. Within the social factors that shape citations' decisions, social proximity is many times more important than the prestige of cited authors. While the cumulative properties of prestige make it a relevant factor to understand the shape of the distribution of citations, there are other social aspects that are more important to understand why a paper gets cited. These results give a new dimension for the analysis of inequalities in academia. Non-meritocratic factors are not only based on the Matthew effect. Working on mainstream topics will imply a higher similarity with a number of potential citing articles. Having a strong network of collaborators will also increase the chances of being cited. Only after considering these two elements is that the Matthew effect becomes relevant to explain most of the citations.

In other terms, while prestige is key for understanding highly cited papers, our results indicate that the reasons behind most citations of non-highly cited papers might be beyond prestige, and closer to semantic and social proximity, among other factors. Shifting the focus from highly cited articles towards the bulk of the distribution of citations creates a space of inquiry that can explain the phenomena that affect most scientists. Differences in impact at the middle of the distribution can have a heavy influence in career development for the majority of scientists, which can in turn affect what science is produced.

These results are limited to economics in the US and differences by field can be observed (see Figs. S2-4). This project opens several avenues for future research. Combining collaboration and citation networks with textual analysis offers opportunities to uncover new insights and raise further questions within the field of the science of science. Extending the analysis to other disciplines and countries, incorporating dynamic analyses, and exploring additional explanatory factors—such as authors' identities or institutional affiliations—could provide a deeper understanding and enhance the findings presented in this study.

# Data & Methods

## Dataset

Data for this article were retrieved from the Web of Science for the period 2008-2023. Given the computational cost of working with the full bibliometric network of citations and authors, we had to restrict the analysis to a specific field. We aimed at building a self-contained collaboration network, and focusing on a specific field can lead to completeness problems, especially for interdisciplinarity research. To mitigate these effects, we focused the analysis on the field of economics, which has a very low degree of interdisciplinarity (Truc et al., 2023), and for which social stratification has been previously studied (Fourcade et al., 2015; Siler et al., 2022). To build our corpus, we use an author-based approach, and use the CWTS author disambiguation algorithm to track authors across publications (D'Angelo & van Eck, 2020). We define the population of focal authors to those that contributed to at least three papers and who published at least 50% of their works within the field of Economics—as defined by the WoS. We include all the publications of the focal authors, and then expand our corpus to include all collaborators of these focal authors and their publications, a necessary step for the completeness of the collaboration network. In the appendix, we also show the results for other twelve fields that, as well as Economics, had more than 50% of their citations generated within their field, as an approximation of closeness (Earth & planetary Science, Philosophy, Orthopedics, Probability & Statistics, Meteorology & Atmospheric Science, Economics, Law, Ophthalmology, Dentistry, General Mathematics, Management, Nuclear & Particle Physics, and Astronomy & Astrophysics).

Nevertheless, many of these fields are rather small and most of their authors also publish outside the field, which can create misleading results as the network is incomplete.

Our dataset includes 12,214 focal authors and all their publications, and when co-authors are considered for distance computations, it expands to 43,467 authors and 71,357 articles, of which 69,081 (97%) include abstracts.

We use the full dataset to compute social distances in the network, but focus on three subsets of data for the analysis. First, we consider all pairs of documents with both an abstract and a citation link in order to compute semantic similarity (*citing-cited pairs, n = 264,436*). Second, we contrast this with a random sample of documents where both have abstract but with no citation link, with the condition that the potentially cited document is older than the potentially citing paper (*non citing document pairs, n = 257,199*). Finally, for social distance, we consider all possible combinations of documents. For a set of *N*=71,357 nodes, there are $N(N-1)/2 = 2,545,875,046$ potential pairs, however, only 35% (*n = 882,777,301*) of these pairs can be observed belonging to the same network component, and therefore have a non-infinite distance. We restrict the rest of the analysis—namely the semantic similarity and the model—to the sample of citing-cited pairs and the negative sample of non citing document pairs, given the computational complexity, and to create a meaningful sample for the model.

## Methods

Our goal is to understand how social proximity, semantic similarity, and prestige affect citing behaviour. We use document pairs as our unit of analysis, and define and operationalize these three concepts as follows.

Social proximity is defined as the degrees of separation between authors. We operationalize this concept using the collaboration network, as authors develop knowledge of and trust in their colleagues' work through collaboration. Direct collaborator pairs are closer in social proximity than authors at a higher degree of separation. This operationalization is limited as there are other ways of social proximity, such as working at the same institution, networking interactions on conferences or social events, or ties outside academia, that are not considered. Self-citations, within this context, are an extreme case in terms of proximity, and can be naturally represented in the collaboration network as a distance of zero. Degrees of separation are defined for any pair of authors. Since our unit of analysis is pairs of documents, we define the distance between two documents as the minimum distance between any pair of authors of each of the documents, or the distance between two documents *A* and *B* is $D(A,B) = \min(d(a\_i, b\_j))$, where $d(a\_i, b\_j)$ is the distance between author *i* of document *A* and author *j* of document *B*.

Semantic similarity is used as a proxy for content similarity between two documents. We compile articles' titles and abstracts to create their embedding representation. This means, the article is projected as a point in a geometrical space. Unlike simple keyword matching, embeddings capture context and meaning, so if two articles use different words that refer to similar meanings— such as *investment*, *capital allocation*, or *asset management*—they will be relatively close in the embedding space. We use the state of the art "multilingual-e5-large-instruct" pre-trained language model (Wang et al., 2024) as a zero-shot classifier to build an embedding of 1024 dimensions. We then apply the cosine similarity between pairs of articles to find how close they are in this embedding space. Cosine similarity calculates the angle between two vectors rather than their raw distance, which is useful because points in the embedding have different magnitudes, but their relative orientation in space matters more for similarity.

We define prestige as the total number of citations accumulated throughout authors' entire career. As in the case of social proximity, since our unit of analysis is document pairs, we consider the author with the highest number of citations as a prestige symbol for any given document (Siler et al., 2022). The argument here is that the highest cited author is in general the most likely to attract prestige-based citations. In this case, nevertheless, the relation between documents is not symmetrical, so we define prestige based on the authors in the cited article, as our research question focuses on the role of prestige in the decision to cite an article.

To compare the differential effect of these three factors, we build a logistic model that predicts the probability of a pair of documents having a citation link given their social and semantic proximity, and the prestige of the authors of the potentially cited document. The degrees of separation in the collaboration network are considered as a series of dummy variables from distance 0 –self-citation– to "6 or more" (see appendix Fig. S5). This rescaling of distance allows us to equalize unconnected documents to the furthest distance in the network, instead of discarding these cases. Interaction terms are included as we expect that for further distances, content similarity and prestige will become more relevant. Finally, as we are not considering all factors that affect citations, such as quality, gender (Nielsen, 2016), or institutional belonging (Kozlowski et al., 2024), we include fixed effects at the cited article level to control for these and other potential factors. We also normalize the independent variables to a 0-1 range to simplify interpretation. The model is defined as follows:

$$\text{logit}(P(y=1)) = \beta_0 + \beta_1 \text{Similarity} + \sum_{i=0}^{5} \beta_{i+2} \text{Step}_i + \beta_8 \text{Prestige}$$
$$+ \sum_{i=0}^{5} \beta_{i+9}(\text{Similarity} \times \text{Step}_i) + \beta_{15}(\text{Similarity} \times \text{Prestige})$$
$$+ \sum_{i=0}^{5} \beta_{i+16}(\text{Step}_i \times \text{Prestige}) + (1|\text{article}) + \epsilon$$

Where Similarity is the semantic similarity, $Step_i$ is a dummy variable for distance *i*, with the reference category being 6 or more degrees of separation. Prestige is the maximum number of accumulated citations by an author of the cited paper.

We use Average Marginal Effects (AME) of each of the independent variables to understand their effect on citations, which can be understood as the expected variation in the probability of a citation for a one-unit change in the covariate, holding all other variables constant at their mean. To illustrate the interactions between independent variables, we created a grid of values representing different combinations of covariate pairs, while keeping all other variables fixed at their means. We then used this grid to predict the model's outcome, providing a clear visualization of how the covariates interact.

# Appendix

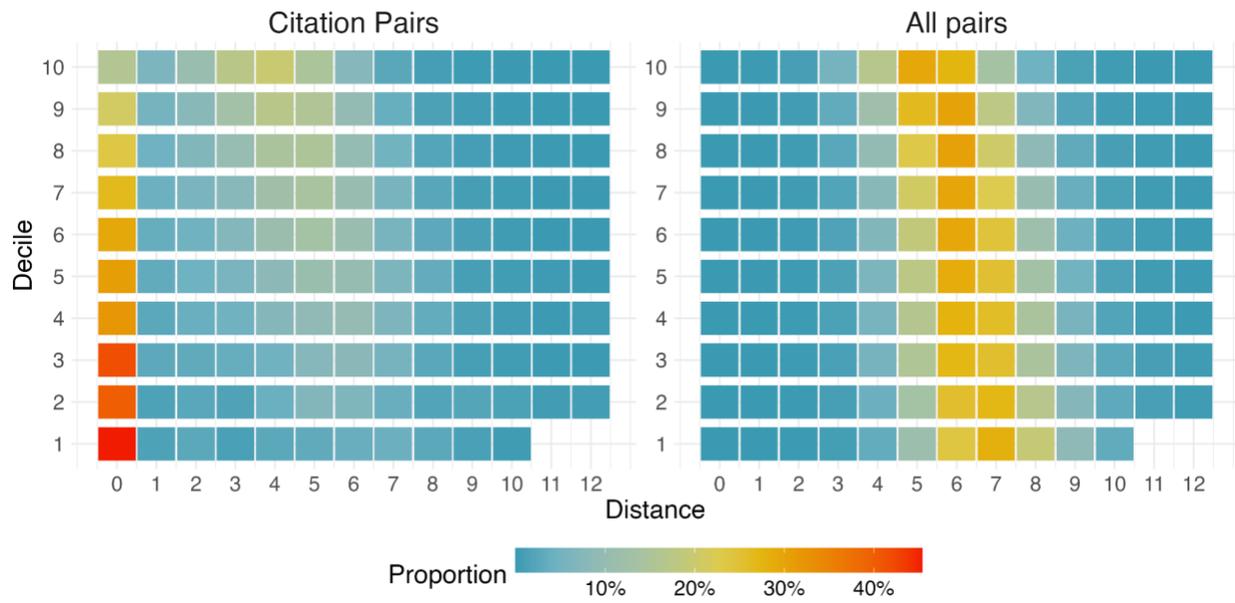

Figure S1. Distribution of distance between articles by deciles of authors' citations on citing-cited pairs and all pairs of articles.

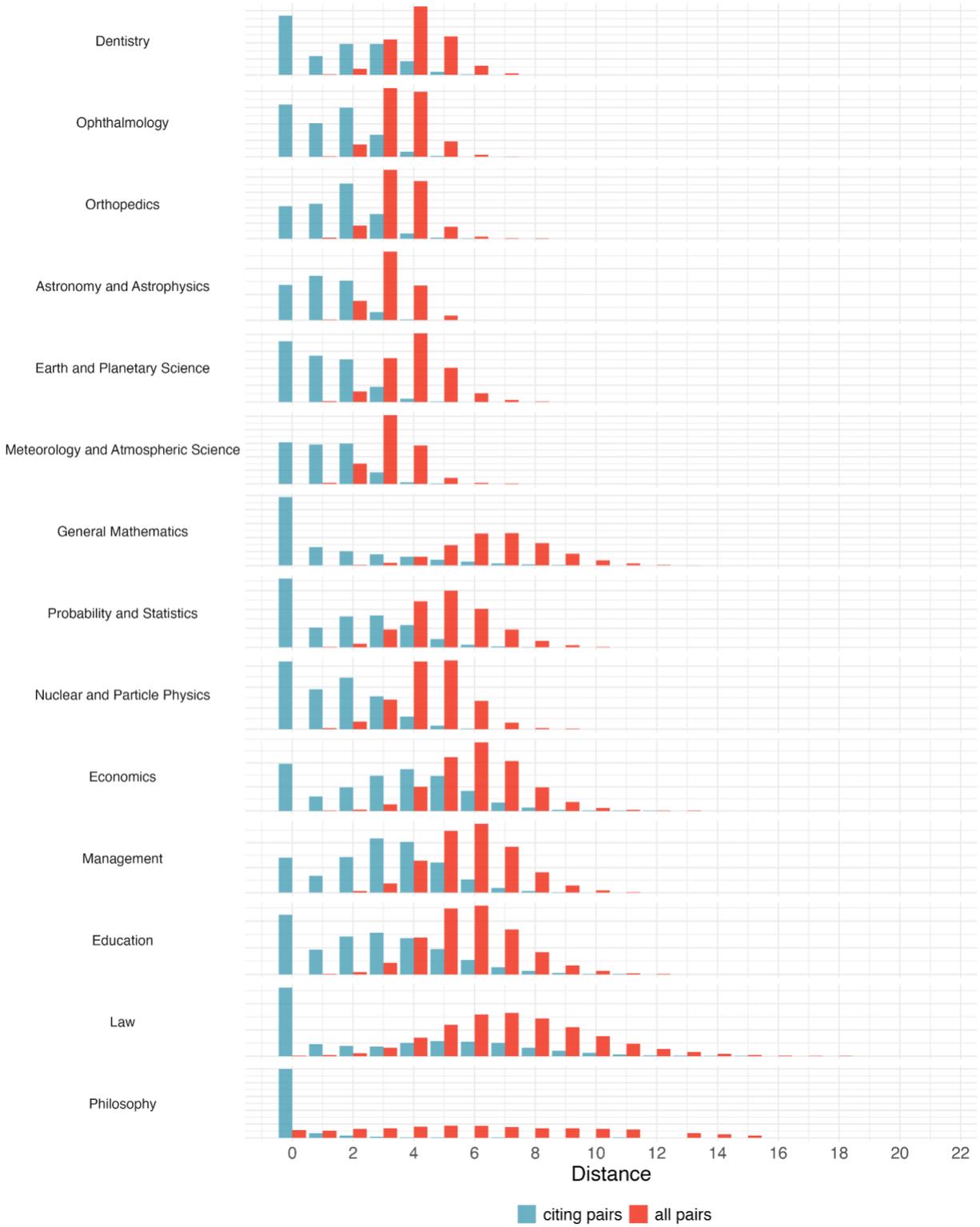

Figure S2. Distribution of distances over different fields

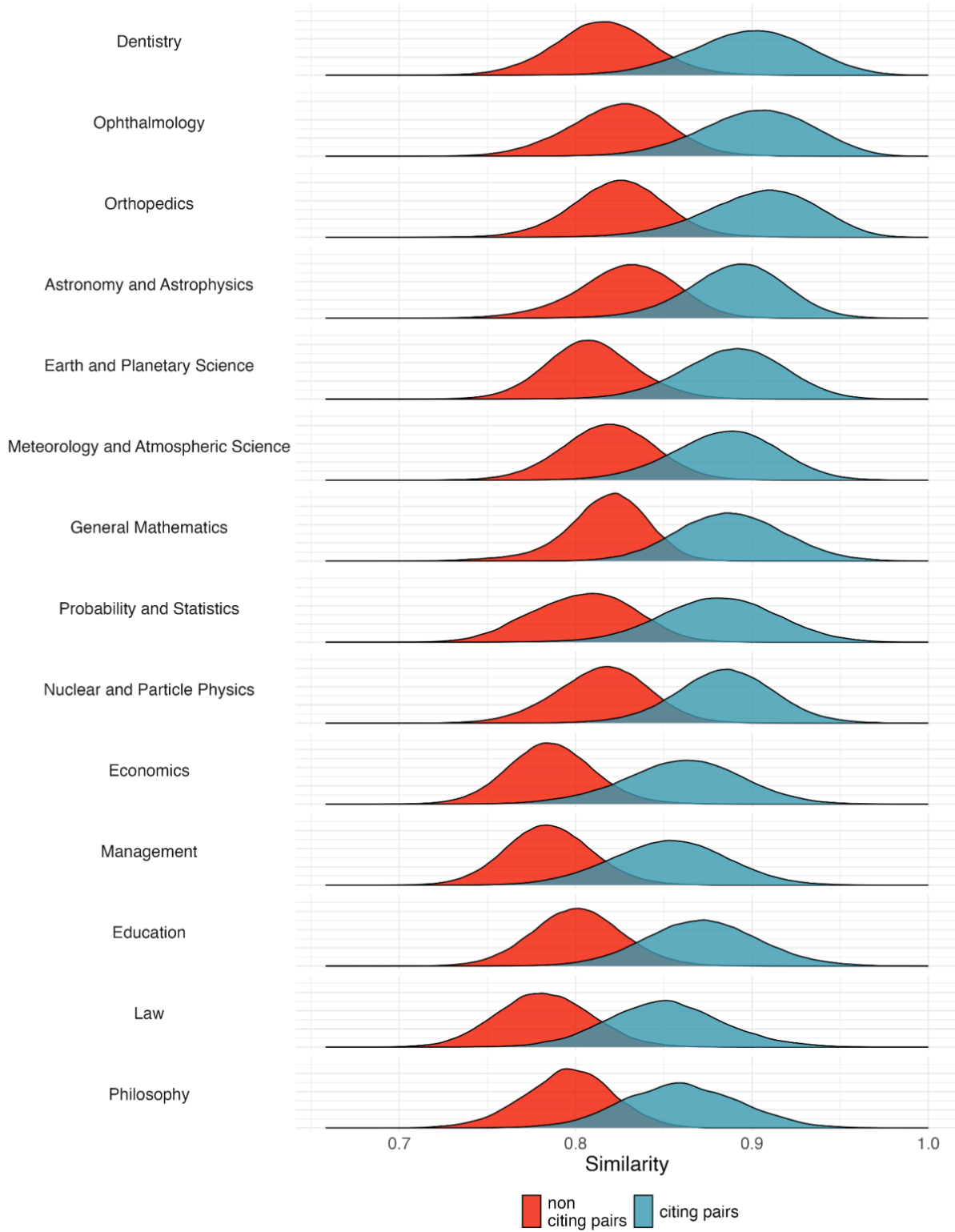

Figure S3. Distribution of semantic similarity over different fields

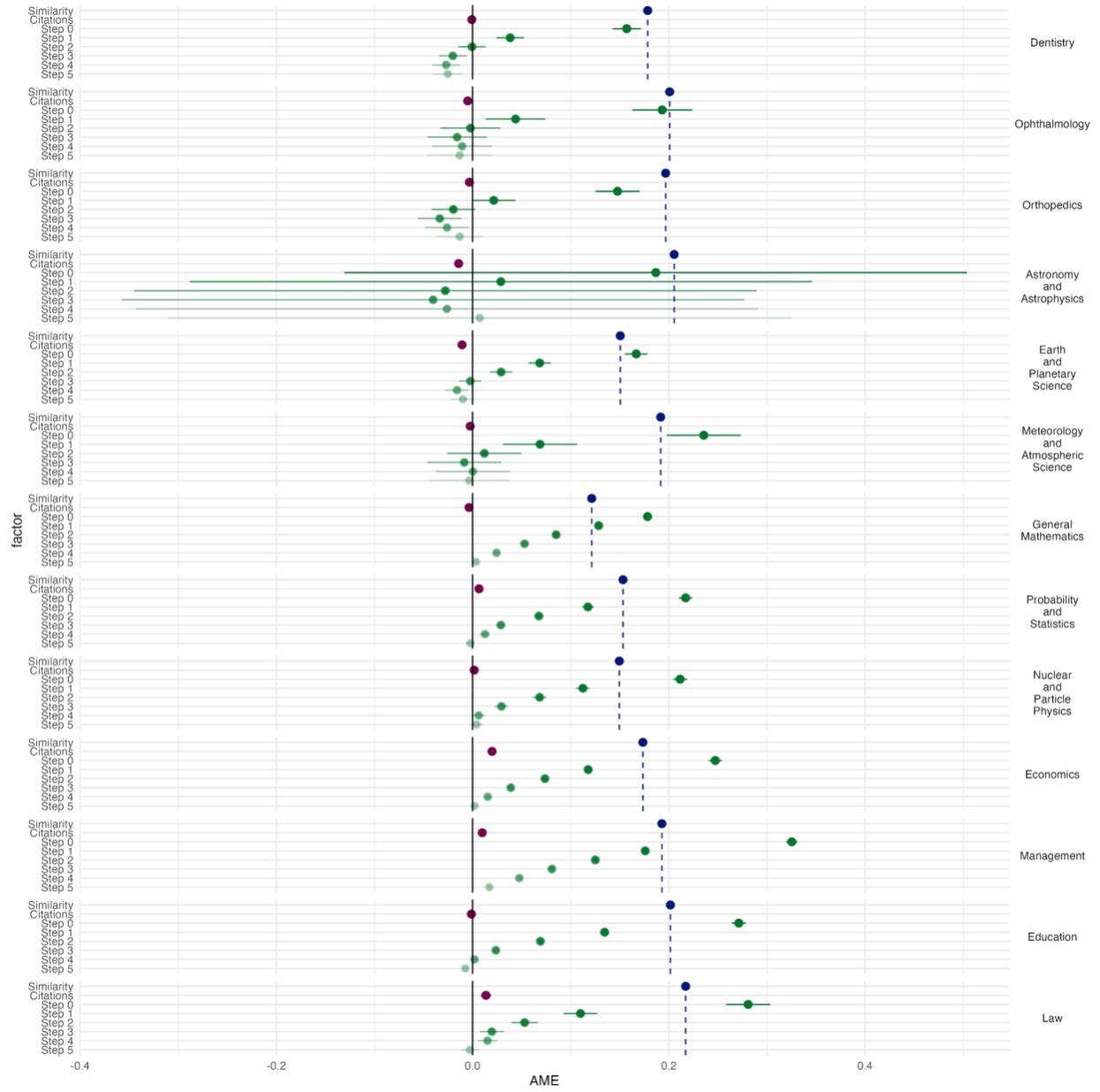

Figure S4. Average Marginal effects on different fields

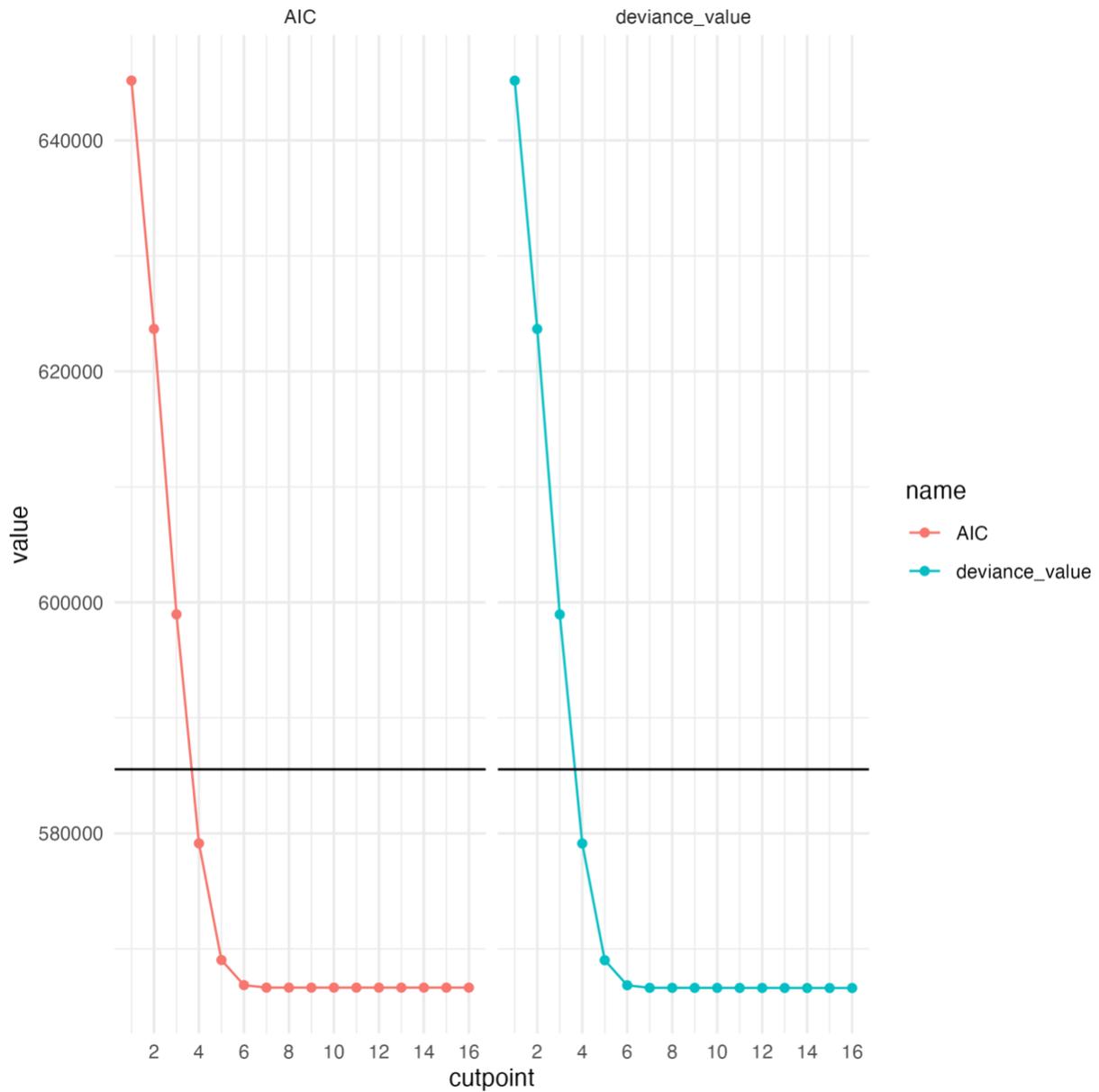

Figure S5. Performance metrics of models including different cut-points of degrees of separation. Each cut-point represents a model with that number of dummy variables, where the last category also includes all further distances and is the reference value. The horizontal line represents the model with distance as a continuous variable. Models including only a self-citation flag (cut-point 1), and co-authors flag (cut-point 2) underperform with respect to the continuous model, but models including up to 4-6 degrees of separation show an improvement with respect to the continuous version. After 6, the improvement of the models are marginal.